\documentclass[10pt]{iopart}
\usepackage{graphicx}
\usepackage{cite}
\usepackage{url}
\usepackage{iopams}

\begin{document}

\title[]{Versatile surface ion trap with fork junction for effective cooling}

\author{Xinfang Zhang$^{1,2}$, Baoquan Ou$^{1,2}$,  Ting Chen$^{1,2}$, Yi Xie$^{1,2}$, Wei Wu$^{1,2*}$ and Pingxing Chen$^{1,2\dag}$}

\address{$^{1}$Department of Physics, College of Liberal Arts and Sciences, National University of Defense Technology, Changsha 410073, China}
\address{$^{2}$Interdisciplinary Center for Quantum Information, National University of Defense Technology, Changsha 410073, China}
\ead{weiwu@nudt.edu.cn, pxchen@nudt.edu.cn}
\vspace{10pt}
\begin{indented}
\item[]July 2019
\end{indented}

\begin{abstract}
Scaling up and effective cooling of ions in surface ion trap are central challenges in quantum computing and quantum simulation with trapped ions. In this theoretical study, we propose a versatile surface ion trap. In the manipulation zone of our trap, a symmetric seven-wire geometry enables innate principle-axes rotation of two parallel linear ion chains, which facilitates the cooling of ions along all principle trap axes. To alleviate contaminating the manipulation zone during ion loading, a symmetric five-wire geometry is designed as the loading zone. And a ``fork junction" connects the loading and manipulation zones, which also enables the shuttling and reordering of ions. A multi-objective optimization procedure suitable for arbitrary junction designs is described in detail, and we present the corresponding optimal results for the key components of our trap. The proposed versatile trap can be used in the construction of large-scale ion quantum processors. The trap also can be used as the multi-ion-mixer or the efficient ion beam splitter, which has the potential applications in quantum simulation and quantum computing, the research of 2D dimensional ion crystals and the guides of quantum microscope, like an electron beam splitter used for quantum matter-wave optics experiments. Interesting topics involving the spin-spin interactions between two ion chains can also be simulated in our trap.
\end{abstract}

\vspace{2pc}
\noindent{Keywords}: surface ion trap, principle-axes rotation, fork junction,

\quad \quad \quad quantum computing and quantum simulation

\maketitle

\ioptwocol

\section{Introduction}

\quad \quad The trapped ions system is one of the most attractive candidates for quantum computing\cite{Klimov2003Qutrit}, offering a long coherence time and a high-fidelity quantum operation\cite{ballance2016high}. The quantum
charge-coupled device (QCCD)\cite{kielpinski2002architecture} is one of the major schemes for the scalability of ion-trap-based quantum information processing\cite{Monroe2013Scaling}. In this scheme, the surface ion trap plays an important role since it completely eliminate the difficulties in the assembly of macroscopic devices such as blade traps and four-rod traps\cite{N2000Investigating,cha2000interface}. The manufacturing of surface ion trap is much simpler\cite{sterling2014fabrication,Chiaverini2005Surface,Stahl2005A,hellwig2010fabrication} using the micro/nanofabrication technologies with recent processes. The trapping performance depends entirely on the electrodes' geometry in the surface ion trap. The symmetric five-wire (FW) geometry trap is widely used in trapped ion scaling\cite{Wright2013Reliable,moehring2011design,Mokhberi2017Optimised}, due to the simplicity of its geometry and fabrication process.

However, with the symmetric FW trap, it's impossible to realize effective ion cooling in the direction perpendicular to the trap surface, because the wavevectors of the cooling laser beams have to be parallel to the trap surface. Fortunately, the effective ion cooling can be achieved by rotating the principal axes of the trap or offering the optical access through a hole in the substrate. However, it is difficult to etch a hole in substrate materials such as silica or sapphire. The rotation of the principal axes is easier to realize effective ion cooling, which is widely used in surface ion trap. The principal axes rotation can be achieved by an elaborate design of the radio frequency (RF) electrodes\cite{Stick2010Demonstration,Allcock2010Implementation,Doret2012Controlling,Amini2009Scalable,Wright2013Reliable,Leibrandt2009Demonstration,Allcock2013A,Niedermayr2014Cryogenic}. Currently, there are several designs to rotate the principle axes of the trap, which has been realized in many trap geometries: the four-wire geometry\cite{Britton2006A,Seidelin2006Microfabricated}, the asymmetric geometry\cite{Narayanan2011Electric,Daniilidis2011Fabrication,Niedermayr2014Cryogenic,Labaziewicz2008Suppression}, the six-wire geometry\cite{Allcock2010Implementation,Stick2010Demonstration} and so on. However, with the four-wire geometry or six-wire geometry (different voltages are applied on two center DC electrodes), the trapped ions above the electrode gap suffer from the gap potential\cite{Schmied2010electrostatics} and the stray charge on the substrate\cite{Chiaverini2005Surface}, which will influence the trapping stability. Besides, it is difficult to realize the segmentation control of the DC electrodes\cite{Chiaverini2005Surface} for four-wire trap. The asymmetric design leads to more capacitive coupling and higher RF loss\cite{Allcock2010Implementation}. Furthermore, the downsides of these geometries will be aggravated when we scale up trapped ion systems. For example, the asymmetric geometry trap suffers from high RF losses, which are proportional to the scale and the width ratio of two RF electrodes\cite{Amini2009Scalable}. For other designs, special designs with additional electrodes are required\cite{Bermudez2017Assessing} to rotate the principle axes at an appropriate angle. These designs have an extremely complex and difficult fabrication process, which involved with multi-layer structures and buried wire technology.

To avoid the problems mentioned above, we propose a novel surface ion trap that has innately rotated principle axes without excess electrodes or asymmetric geometry and enables 2D large-scale parallel ion chains trapping. The chip consists of a seven wire (SW) geometry trap, a symmetric FW geometry trap, and a fork junction. The SW trap generates double wells simultaneously at a distance of 200 $\mu$m and the rotation of their principal axes at $\pm$35 degrees. This not only enables trapping two parallel ion chains, but also can effectively cool ions along all principle axes in the manipulation zone. We design a symmetric FW trap as the loading zone, which produces a single well located 100 $\mu$m above the trap surface. A fork junction is designed to shuttle ions from the loading zone to the manipulation zone after pre-cooling. The shuttling path is split at the transportation zone forming a ``fork junction". The junction geometry is optimized by a combination of the ant colony algorithm and a multi-objective function.

Our design can be applied in many interesting research fields. In the manipulation zone, two species of ions can be trapped in the double wells independently. After the ordered transport of ions from double wells to the symmetric FW trap, the mixed and ordered two-specie ion chains can be obtained. So, the trap can be used as a mixer of ions\cite{Bermudez2017Assessing}, which can offer a flexible scheme for sympathetic cooling ions\cite{larson1986sympathetic}. In addition, we can split an ion chain in the loading zone into two ion chains trapped in the double wells. The split ability (like an ion beam splitter) should be useful in the studies of 2D dimensional ion crystals\cite{britton2012engineered,porras2006quantum} and in the industrial applications as a guide of quantum microscope such as the ion etching technology\cite{tachi1988low} and the electronic imaging technology\cite{Hoffrogge2011Microwave,Jakob2015Microwave,hammer2015microwave}. The effective spin-spin interaction between two ion chains can also be studied based on our trap\cite{wilson2014tunable,welzel2011designing}.

\section{Design of linear electrode zones}

\quad \quad A symmetric FW trap has been designed, fabricated and tested in our lab\cite{Xie2017Creating,Ou2016Optimization}. To further improve the confinement performance, realize effective cooling of ions, and extend the application, a versatile surface ion trap is designed as shown in Figure 1. There are three parts in the trap: two linear electrode zones of the loading zone and the manipulation zone, and a fork junction. The symmetry FW trap serves as the loading zone where ions are trapped at the height of $h = 100 \mu$m above the surface, and ions can be transported through the junction into the SW trap (the distance of double wells d is $200 \mu$m). The advantage of our design is that it provides the chance of effective ion cooling without redundant (DC or RF) electrodes and realizes trapping two parallel ion chains, which can confine twice as many ions as the symmetric FW trap.

\begin{figure}
\centering
\includegraphics[width=8cm]{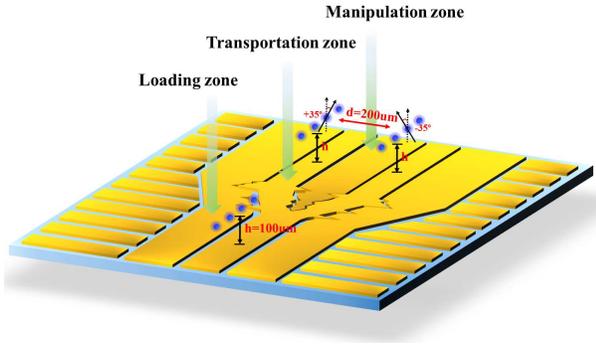}
 \caption{\label{vid:PRSTPER.4.010101}%
 The trap with different quantum zones, including a SW trap as the manipulation zone to generate double wells with rotating principle axes for effective cooling of ions, a symmetry FW trap as the loading zone to load and pre-cool ions, and a fork junction as the transportation zone to shuttle ions.
 }%
\end{figure}

\subsection{Design method}

\quad \quad The surface ion trap has two main components: the metal electrodes (DC and RF) and the insulating substrate\cite{Seidelin2006Microfabricated,Britton2006A}. When RF voltages are applied to the electrodes, RF losses are inevitably generated on the dielectric substrate. Large RF voltages are usually applied on the RF electrodes to constrain ions in the strong-binding regime, but a higher RF voltage generally produces a greater loss on the substrate\cite{lee2003design}. In general, the RF loss on the dielectric substrate is mostly transformed into thermal energy, which results in a temperature rise of the trap that significantly influences the trapping stability. One simple way to remedy this problem is to reduce the RF voltage as much as possible when the ions are always trapped in the strong-binding regime.

Our design method is different from general designs\cite{Stick2010Demonstration,Allcock2010Implementation,Doret2012Controlling,Amini2009Scalable,Wright2013Reliable,Leibrandt2009Demonstration,Allcock2013A,Niedermayr2014Cryogenic}. Instead of directly optimizing the trapping height, depth, frequency, and $q$ parameter of the Mathieu differential equation\cite{Leibfried2003Quantum}, we use the pseudopotential curvature as the optimal parameter\cite{Schmied2009Optimal} to obtain an deep trapping depth and high trapping frequency with the RF source $U_{RF}$. We consider ions of mass $m$ and charge $q$ that are confined at height $y$ by the pseudopotential generated by the RF electric field with amplitude $\vec E(\vec r)$ at angular frequency $\Omega_{RF}$. The pseudopotential is
\begin{equation}\label{1}
\centering
  \qquad \qquad \quad \Phi(\vec r)=\frac{q^{2}\|\vec E(\vec r)\|^{2}}{4m\Omega^{2}_{RF}}.
\end{equation}
At arbitrary positions $\vec r$ above the trap, the pseudopotential curvature tensor is proportional to the square of the electric potential curvature tensor\cite{Schmied2009Optimal}, i.e. $\Phi^{(2)}(\vec r)=\partial_{\alpha}\partial_{\beta}\Phi(\vec r)$. According to Laplace equation, the trace of this tensor is zero, i.e. $Tr\Phi^{2}(\vec r)=0$, and the elements of this tensor are dependent on the trap geometry. In the linear Paul surface ion trap, the pseudopotential curvature tensor matrix can be described as
\begin{center}
\begin{equation}\label{3}
  \qquad \qquad \Phi^{2}(\vec r)=\left(
                     \begin{array}{ccc}
                       1 & 0 & 0 \\
                       0 & -1 & 0  \\
                       0 & 0 & 0 \\
                     \end{array}
                   \right).
\end{equation}
\end{center}
The matrix as the boundary conditions can be used to design the RF electrodes' geometry in terms of the spatial position of ions. In this process, the pseudopotential curvature tensor is applied globally to all ions. According to Ref. \cite{Schmied2009Optimal}, the dimensionless curvature $\kappa$ can quantify the trap strength as the following expression
\begin{equation}\label{4}
 \qquad \qquad \kappa=|det\Phi^{(2)}(\vec r)|^{\frac{1}{3}}(\frac{y^{2}}{U_{RF}}).
\end{equation}
Subject to the Mathieu stability requirements, it prefers to optimize the trap geometry so that the parameter $\kappa$ is maximized for the given constraints. Furthermore, the higher trapping frequency can be achieved with the larger $\kappa$, when the same RF voltage is applied. A higher secular frequency is desired because it not only allows tighter confinement, faster ion transportation\cite{Yeo2007On}, more effective cooling, and less sensitivity to external noise\cite{Zhu2006Trapped}, but also can realize multi-qubit entangling gate operations using motional modes\cite{Cirac1995Quantum}.

\subsection{The manipulation zone design}

\quad \quad Based on the above method, we designed and optimized the symmetric SW geometry trap. The RF electrodes' geometry is determined by two parallel linear ion chains. Based on the spatial position of the ions, the electric potential is solved by the boundary element method (BEM)\cite{chen1992boundary} to obtain the electrode geometry for the expected $\kappa$ value. The potential generated by the electrodes is analytically calculated in the gapless plane approximation\cite{Wesenberg2008Electrostatics} using the SurfacePattern software package\cite{SurfacePattern}. Here, the desired trapping height (RF null) is 100 $\mu$m above the trap surface, and the two ion chains are juxtaposed at a distance of 200 $\mu$m. Finally, the RF electrodes of the SW trap include one $RF_{Centre}$ and two $RF_{Side}$, whose widths are 120 $\mu$m and 167 $\mu$m, respectively. Both ground electrodes are 83 $\mu$m width, as shown in Figure 2(a). As we can see from the pseudopotential distribution generated by RF electrodes in Figure 2(b), there is an innate $\pm35$-degree angle between the trap principal axes and the $y$ detection perpendicular to the trap surface. A laser-cooling beam parallel to the trap surface can cool ion along all principle axes, which is expected to achieve effective ion cooling to their vibrational ground state.

\begin{figure}
\centering
\includegraphics[height=5cm]{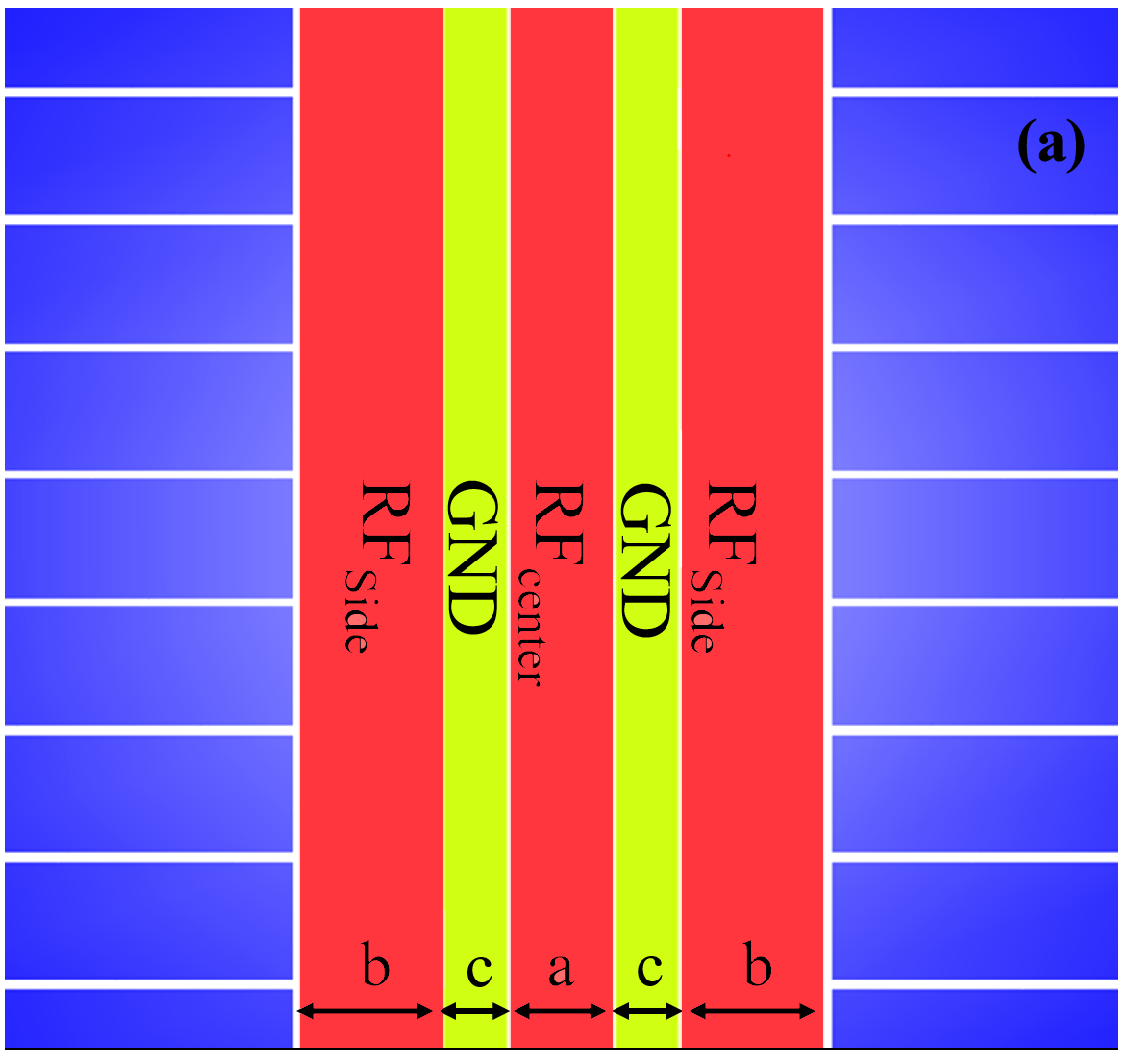}
\quad
\includegraphics[width=5.4cm,height=4.4cm]{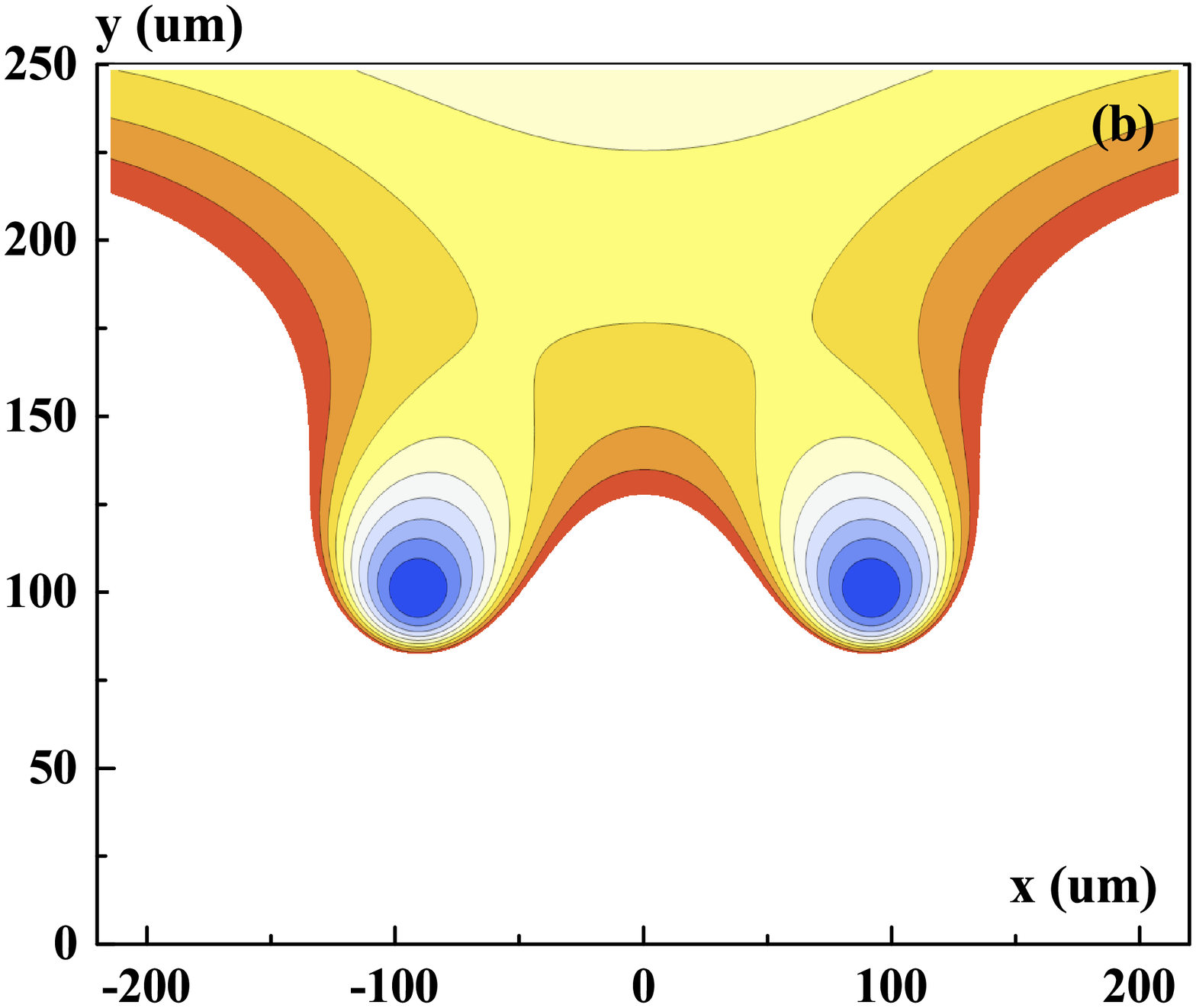}
\includegraphics[width=2.5cm,height=4.1cm]{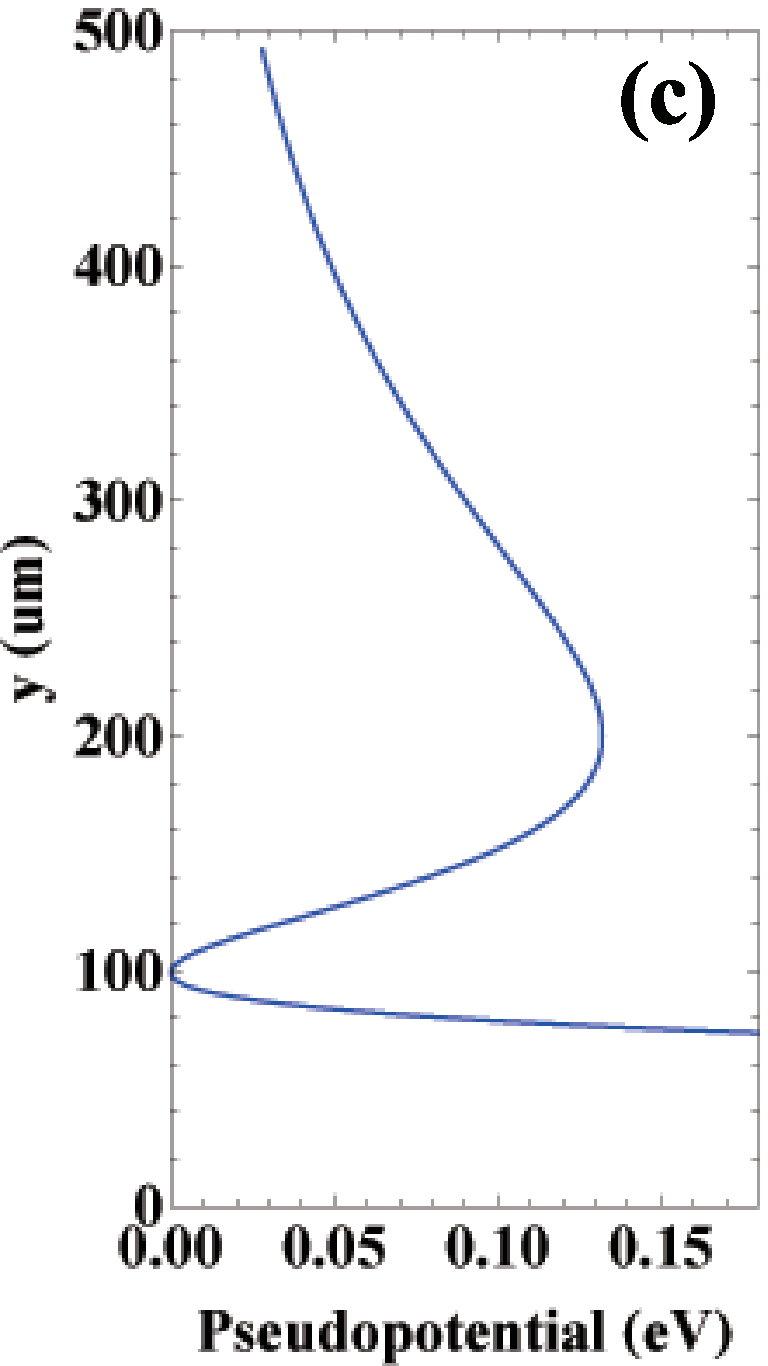}
 \caption{\label{vid:PRSTPER.4.010101}%
Parameters of the SW trap and the pseudopotential distribution.
  (a) Diagram of the trap layout and dimensions. The trap consists of an $RF_{Centre}$ electrode of width a = 120 $\mu$m, a pair of $RF_{Side}$ electrodes of width b = 167 $\mu$m, two ground electrodes of width c = 83 $\mu$m, and the segmented DC control electrodes with 150 $\mu$m width.
  (b) Radial contour plot of total pseudopotential generated by (a). There are double wells in the $x$ direction. The pseudopotential decreases from red to blue.
  (c) 1D plot of the pseudopotential along the line through the null and saddle points. The trap depth is about 0.129 eV, and the trapping height and the escape point are approximately 98 $\mu$m and 201 $\mu$m above the trap surface, respectively.
 }
\end{figure}

The trapping height of the $^{40}Ca^{+}$ ions is about 98 $\mu$m simulated by the BEM with a 100-V-amplitude RF source at a 25 MHz frequency. Figure 2(c) shows that the escape point (saddle point) is approximately 201 $\mu$m, and the trap depth is 0.129 eV, which is five times deeper than that in Ref. \cite{Hong2016Guidelines} (see Table 1) obtained with the 155 V. The secular frequencies in the radial $x$ and $y$ directions are approximately 2.731 MHz and 2.487 MHz, respectively. We can achieve deeper trap depth and higher trap frequency by optimizing the electrode geometry. The distance between two RF nulls is 200 $\mu$m, as shown in Figure 2(b), which can trap two parallel linear ion chains along the RF electrodes. The distance between the RF nulls can be controlled by the RF voltages applied to $RF_{Centre}$ and $RF_{Side}$ electrodes. However, the phase difference between RF electrodes needs to be overcome, since it causes excess micromotion. In addition, the $RF_{Centre}$ electrode requires a DC voltage offset to achieve the stable trapping ions in the spatial overlap between the RF null lines and the minimum of electrostatic potential.

\subsection{The loading zone design}

\quad \quad A separate loading zone can alleviate contamination of the other zones\cite{Daniilidis2011Fabrication}, which can prolong the ions' lifetime. According to recent reports \cite{Sage2012Loading,Boldin2018Measuring,Sedlacek2018Distance,Tanaka2012Micromotion,Wright2013Reliable,moehring2011design,Mokhberi2017Optimised}, it is more convenient to use a symmetric FW trap for the pre-cooling and loading of ions. In addition, the loading zone should be capable of the fast loading and efficient shuttling ions. When ions are transported from the loading zone to the manipulation zone, the ions' parameters (e.g., trapped height, depth and frequency) should remain as constant as possible to ensure that the motional modes of the ions are not excited.

The RF geometry of the symmetric FW trap is solved with a 1D linear ion chain at 100 $\mu$m above the trap surface. The calculation process is the same as for double wells. The achieved trap is composed of two RF electrodes of 130 $\mu$m width and a ground electrode of 108 $\mu$m width, as shown in Figure 3(a). The width ratio between the RF electrode and the ground electrode is about 1.2, which is the same as in Ref. \cite{house2008analytic} for symmetric FW trap. The performance of the trap is simulated using BEM for trapped $^{40}Ca^{+}$ ions with an applied 100 V RF voltage at 25 MHz frequency. The pseudopotential distribution of this design is shown in Figure 3(b). The trapping height and escape point are approximately 102 $\mu$m and 190 $\mu$m above the trap surface, respectively. The trapping depth is approximately 0.122 eV. In the loading zone, a deeper trap depth improves the stability of trapping ions, which can shorten the working time of the atom oven to reduce contamination of the trap surface. In addition, it's better to ensure that there is the same trapping depth in different zones, since we will shuttle ions between different zones. The secular frequencies in the radial $x$ and $y$ directions are 2.845 MHz and 2.822 MHz, respectively. Here, we achieve a higher frequency by using a relatively low voltage for the RF source. The Mathieu parameter $q$ along $x$ is 0.229.
\begin{figure}
\centering
\includegraphics[width=5.1cm,height=5cm]{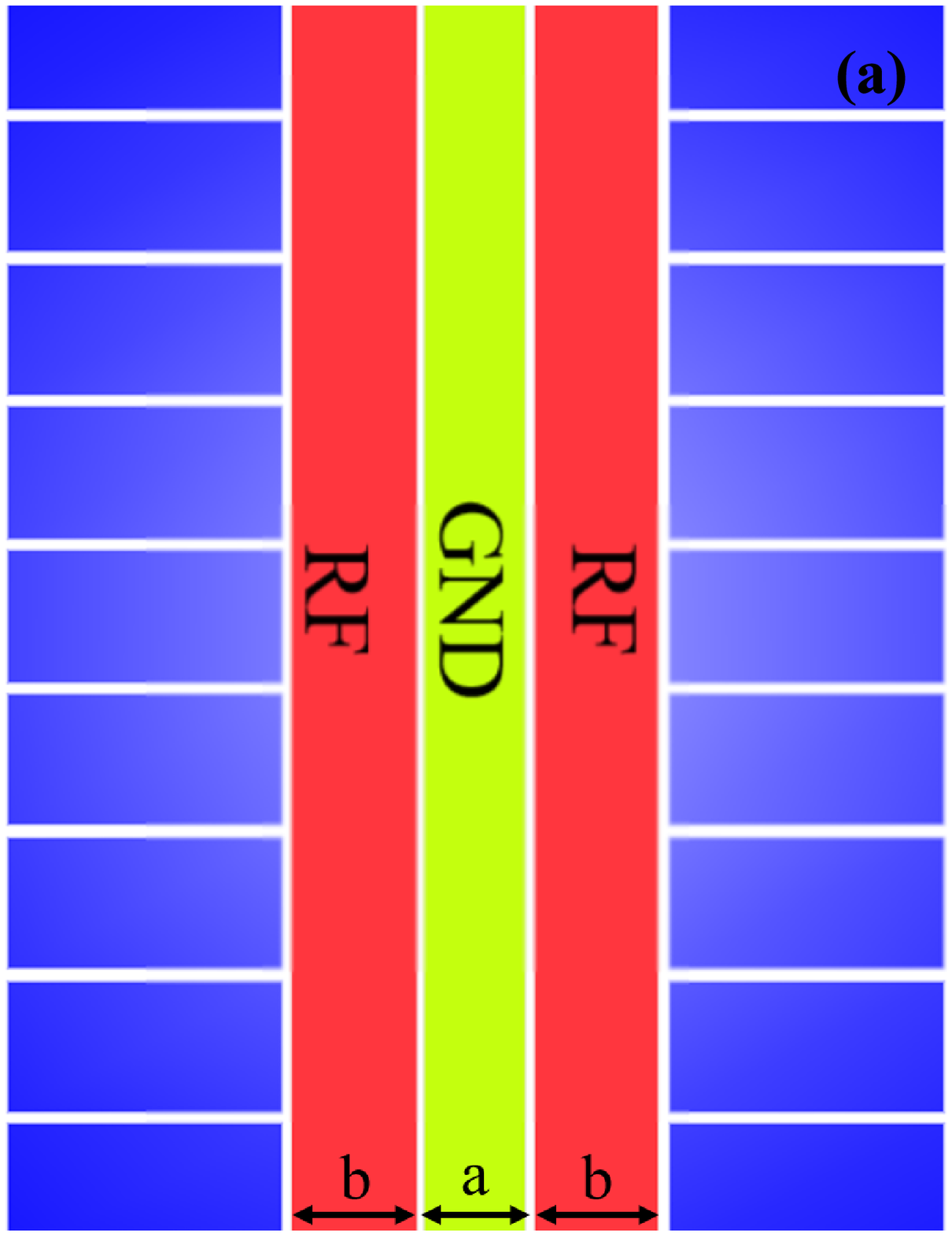}
\quad
\includegraphics[width=5.4cm,height=4.4cm]{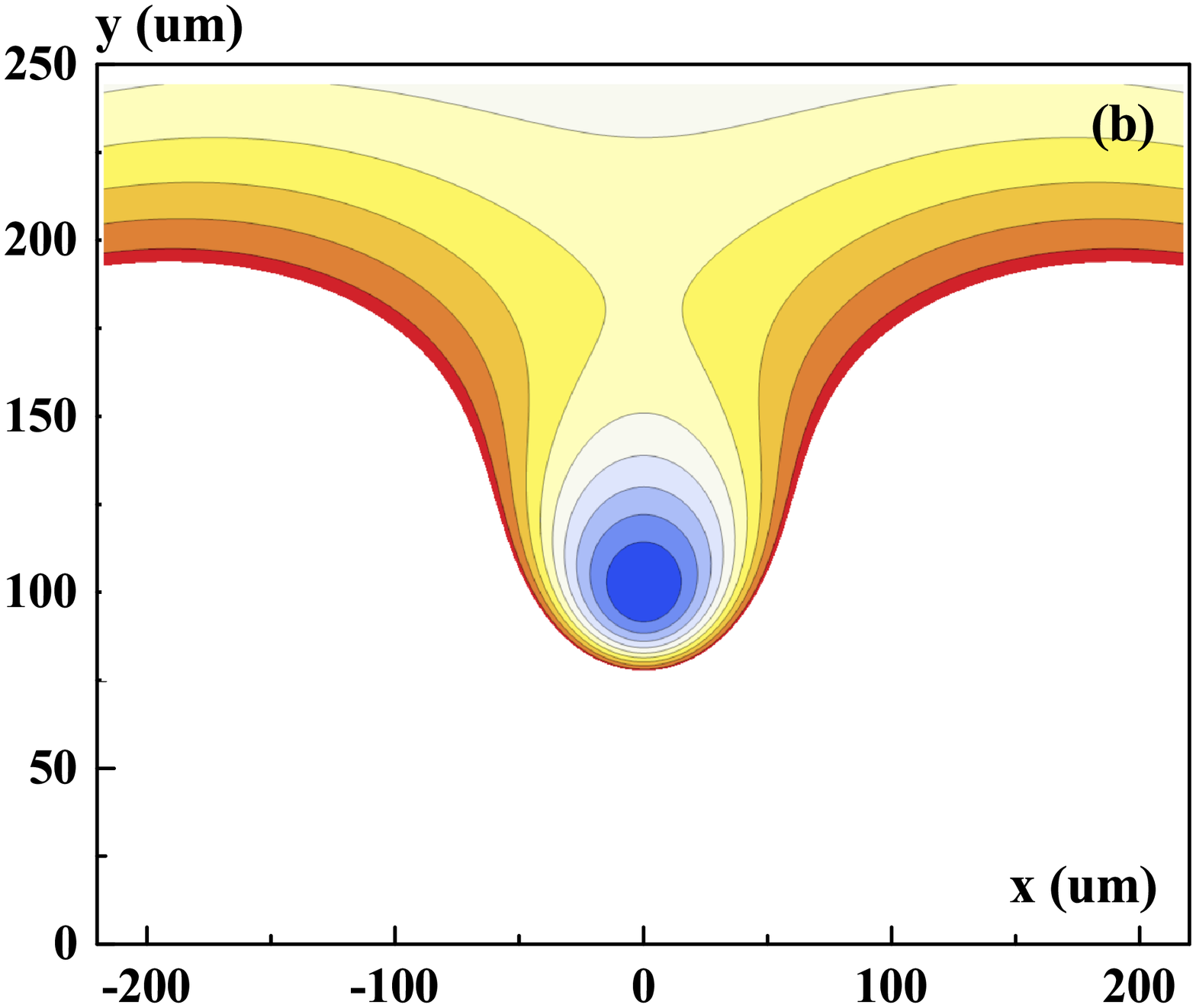}
\includegraphics[width=2.5cm,height=4.1cm]{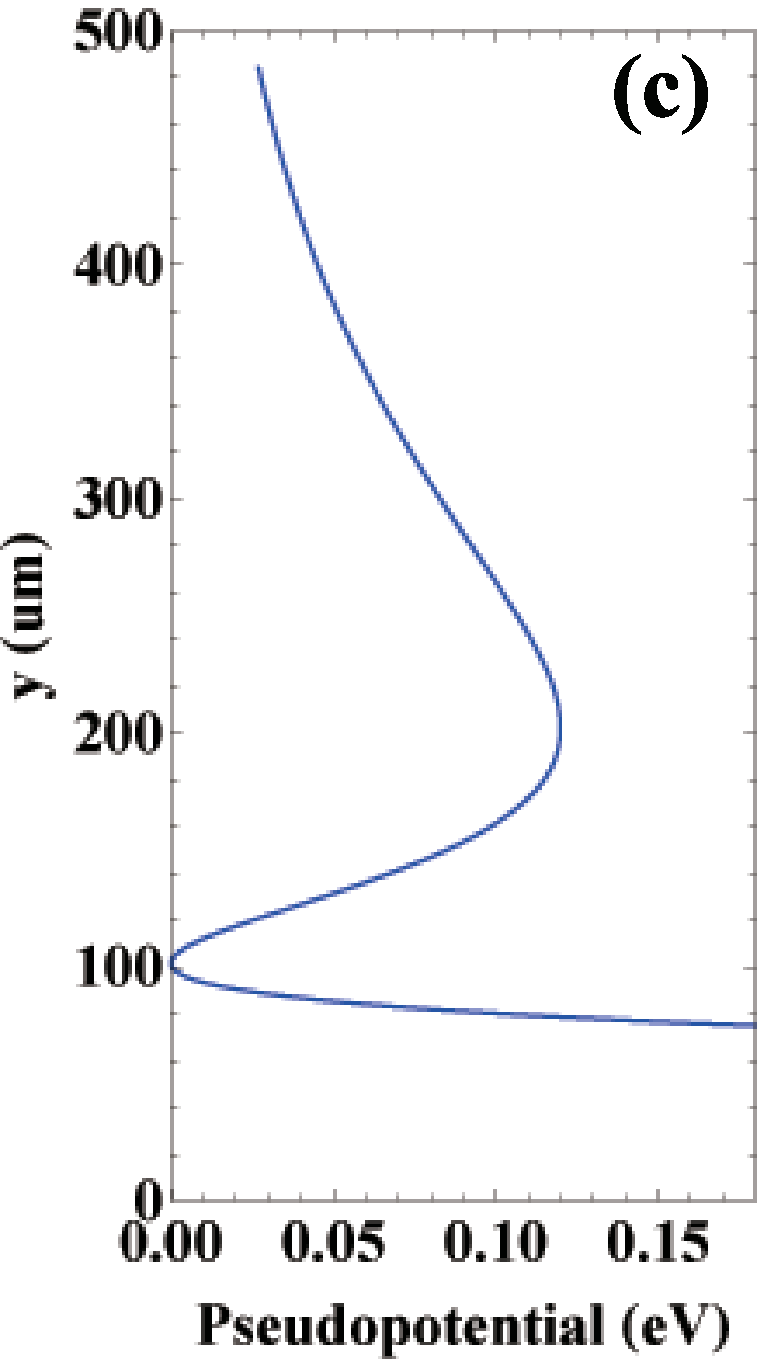}
 \caption{\label{vid:PRSTPER.4.010101}%
Parameters of the FW trap and the pseudopotential distribution.
  (a) Diagram of the trap layout and dimensions. A ground electrode of width a = 108 $\mu$m, a pair of RF electrodes of width b = 130 $\mu$m and the segmented DC control electrodes with 150 $\mu$m width..
  (b) Radial contour plot of total pseudopotential generated by (a). The pseudopotential decreases from red to blue.
  (c) 1D plot of the pseudopotential along the $y$ direction ($x = z = 0$). The trap depth is approximately 0.122 eV, and the trapping height and the escape point are approximately 102 $\mu$m and 190 $\mu$m above the trap surface, respectively.
 }
\end{figure}

\section{Multi-objective optimization of fork junction}

\quad \quad A fork junction is designed to shuttle ions between loading zone and manipulation zone (as described in sections 2.2 and 2.3), which allows orderly merging of two parallel linear ion chains from the double wells to the single well by sequential control of voltages on the DC electrodes, and vice versa. For ions in the near-motional ground states as quantum bits, it is important that shuttling ions between different zones is accomplished with a high success probability and a low motional-energy gain. For this purpose, we use the multi-objective function and the ant colony optimization (ACO) algorithm\cite{Guo2012Ant} to optimize the junction geometry.

\subsection{Analysis of objective functions}

\quad \quad The optimal objectives include equating the minimum values of the pseudopotential, minimizing the pseudopotential gradient, minimizing the trapping height fluctuation, and unifying the shape of the pseudopotential tube (the equipotential surface with a certain radius to the RF null line) along the shuttling path.

The minimum value of the pseudopotential should ideally be equal in the shuttling paths. However, there are some pseudopotential barriers along the shuttling paths (unlike linear electrodes, which generate a uniformly distributed pseudopotential along the RF electrodes) produced by the initial junction electrodes. The pseudopotential barriers on the shuttling paths will reduce the trap depth and weak trapping ability. Eliminating the pseudopotential barriers is the first objective.

In the shuttling process, the heating rate is mainly derived in the pseudopotential gradient, which is the first derivative of the pseudopotential\cite{Blakestad2009High}, since the pseudopotential gradient will result in exciting the motional modes. Therefore, the second objective is to find a junction geometry that maintains a low pseudopotential gradient.

The fluctuation of the trapping height should be as small as possible. The large fluctuation affects the stability and repeatability of the shuttling ions. The minimum potential generated by the time-dependent voltages of the DC electrodes will not overlap with the RF null line at all time, which will result in the aggravated micromotion and heating ions. Furthermore, the trapping height fluctuation lead to misalignment between ions and the laser beam. The third objective should be to ensure a consistent trapping height throughout the shuttling process.

The secular-frequency different between the linear zone and the junction zone will result in changing the energy of ions in the shutting process, which is proportional to the number of shuttling ions. To keep the secular frequency as stable as possible, it is necessary to limit trapping ions above the junction electrodes to be as strong as the symmetric FW and SW traps. Thus, unifying the shape of the pseudopotential tube is the last optimal objective.

\subsection{Multi-objective optimization}

\quad \quad The terms of the multi-objective function are described in Table 1, where $F_{i} (i = 1, 2, 3, 4)$ denotes the different objective functions. The pseudopotential barriers are obtained by finding the pseudopotential extremums along the RF null line $y|_{\Phi_{min}}$ (the $\Phi_{min}$ is the pseudopotential corresponding to the RF null line); $l_{min}$ and $l_{max}$ are the start and the end points of the ion-shuttling path, respectively, and $l$ denotes the whole path. $N_{l}$ and $\Delta$ denote the number of segments on the path and the radius of the pseudopotential tube, which is the distance from the RF null line to the $\Phi_{const}= 10$ meV pseudopotential.
\begin{table*}
\small
\caption{\label{tab:table3}
Mathematical expression and description of the multi-objective function.
\texttt{}}
\begin{tabular}{lp{10.5cm}lp{8cm}}
\br
\multicolumn{1}{l}{Function}&
\multicolumn{1}{l}{\quad Description}\\
\hline

$F_{1}=\int^{l_{max}}_{l_{min}}\Phi(y|_{\Phi_{min}},l)dl$ & The average of the pseudopotential along the trapped path (the pseudopotential minimum).\\
\mr
$ F_{2}=\int^{l_{max}}_{l_{min}}(\frac{\partial|E\cdot\vec l|}{\partial l})dl$ & Average slope of the electric field along the shuttling path (a measure of heating).\\
\mr
$ F_{3}=\int^{l_{max}}_{l_{min}}|y|_{\Phi_{min}}-h|dl$ & Height excursion to $h=100 \mu m$ along the trapped path.\\
\mr
$F_{4}=\sqrt{\frac{1}{N_{l}}\sum_{l}{((y|_{\Phi_{const}}-y|_{\Phi_{min}})-\Delta)^{2}}}$ & The 10 meV contour (a measure of trapping potential shape and secular frequency compared to the linear region).\\
\br
\end{tabular}
\end{table*}
To effectively optimize by the multiple objective functions, we adjust the multi-objective function with the weight factor and the normalization operation:
\begin{equation}\label{5}
  \quad  F_{multi-objective}=\sum^{4}_{i=1}\omega_{i}(\sigma_{i}\cdot F_{i}),\quad \sigma_{i}=\frac{1}{F^{min}_{i}}.
\end{equation}
A small $F_{multi-objective}$ is expected because it indicates better optimization. Where $\omega_{i}$ is the weight factor of objective function $F_{i}$, which determines the contribution of the corresponding objective function in the optimization, and $\sigma_{i}$ is a normalization factor that is equal to the reciprocal of the minimum value obtained when the objective function is independently optimized. The value range of $\omega_{i}$ is between 0 and 1, which is adjusted reasonably according to the objective.

\subsection{Optimization results}

\quad \quad Initially, we simply connected the RF electrodes of the FW trap and the SW trap by the initial junction as shown in Figure 4(a). The electrodes' geometry (gray part) of the SW trap and the SW trap has been maintained based on above optimization. Only the initial geometry of the junction (yellow part) is optimized by the optimization method discussed above. The optimal junction geometry is obtained by adjusting the 38 control points as shown in Figure 4(b), where the $P1 \ldots P19$ points and the $P20 \ldots P38$ points are on both side edges of ground electrode. Thus, we have 38 degrees of freedom, which can only be adjusted in the $x$ direction within the yellow area. The locations of these points are modified using an ACO algorithm and multi-objective functions.
\begin{figure}
\includegraphics[width=4cm,height=3.8cm]{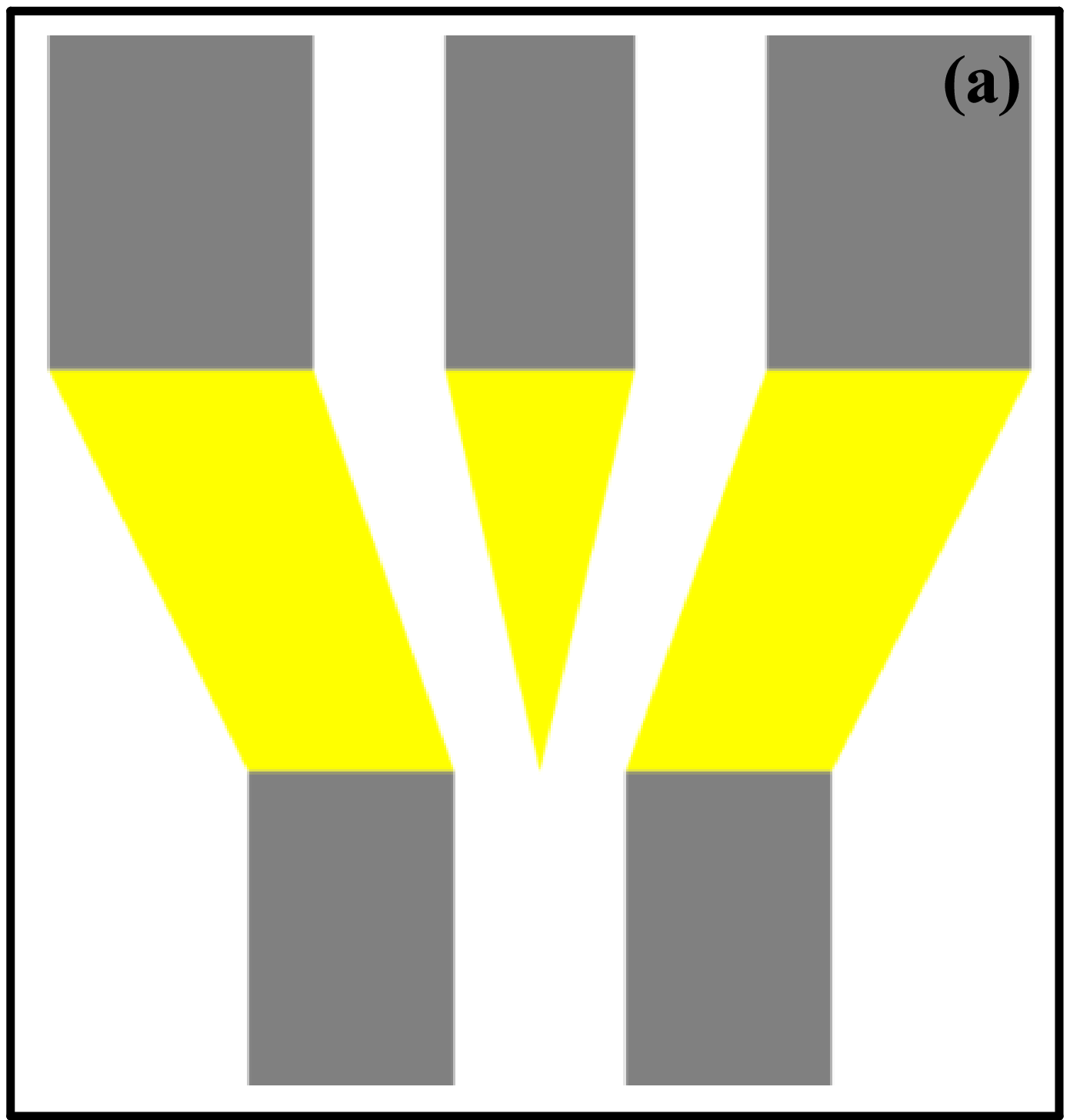}%
\includegraphics[width=4cm,height=3.8cm]{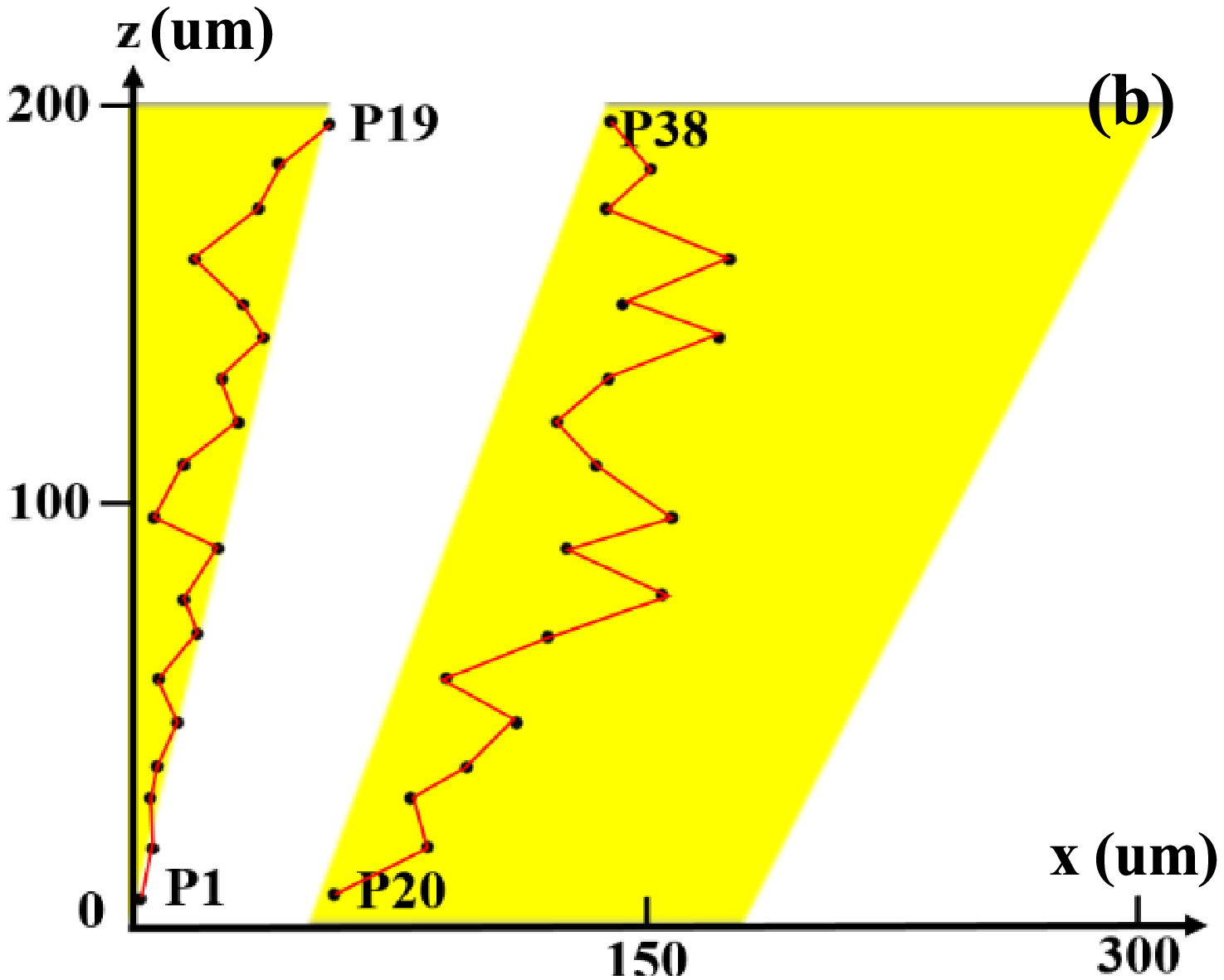}
 \caption{\label{vid:PRSTPER.4.010101}%
Diagram of the local electrodes.
  (a) Layout of symmetric FW and SW traps (gray) and initial geometry of the junction (yellow).
  (b) The half of the optimized layout of the junction, since the junction is symmetrical. The 38 dots denote the variables, which move independently in the $x$ direction
 }%

\end{figure}

\begin{figure*}
\centering
\includegraphics[width=7.45cm,height=6.3cm]{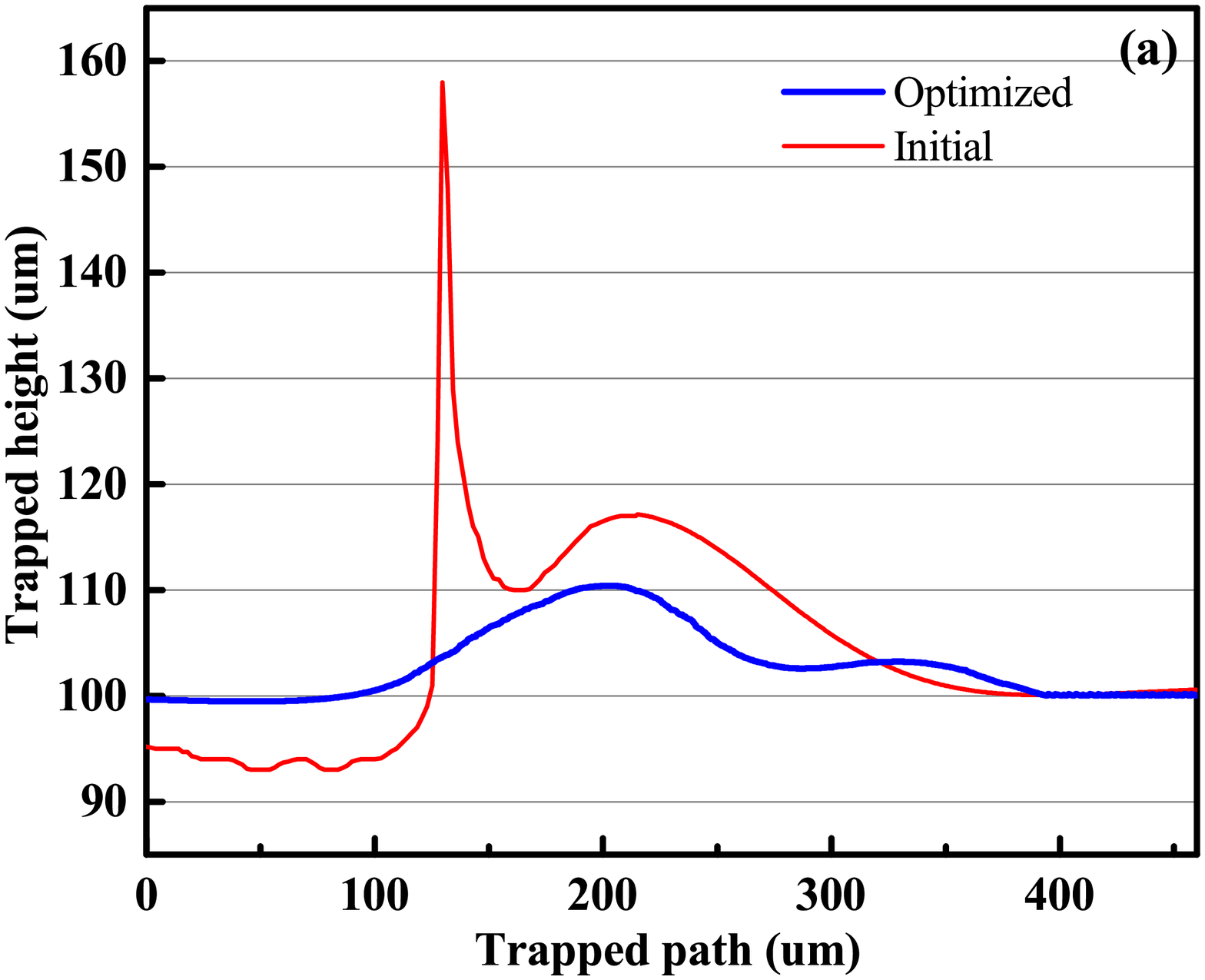}
\quad
\includegraphics[width=7.4cm,height=6.5cm]{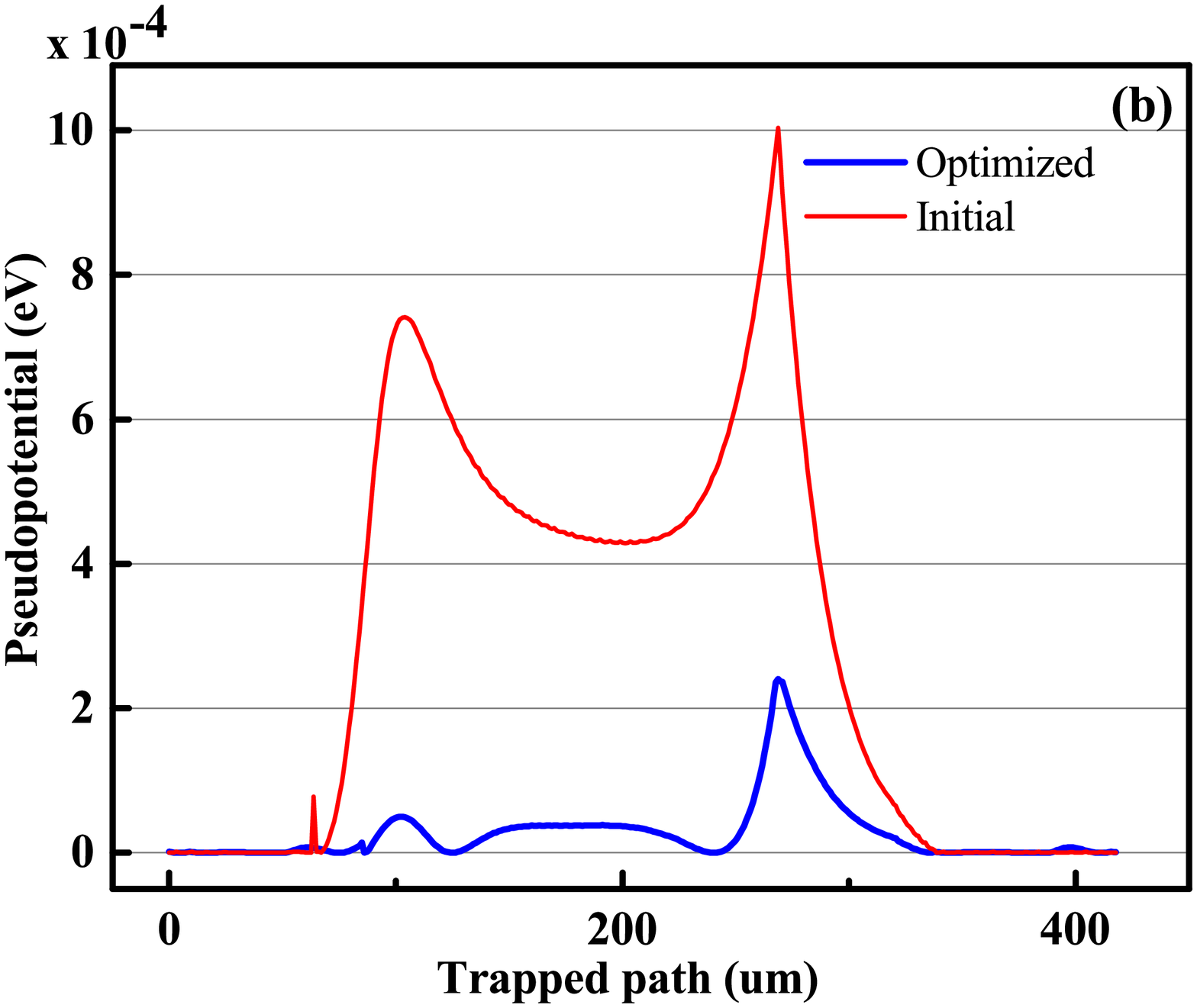}\quad

\includegraphics[width=7.25cm,height=6.45cm]{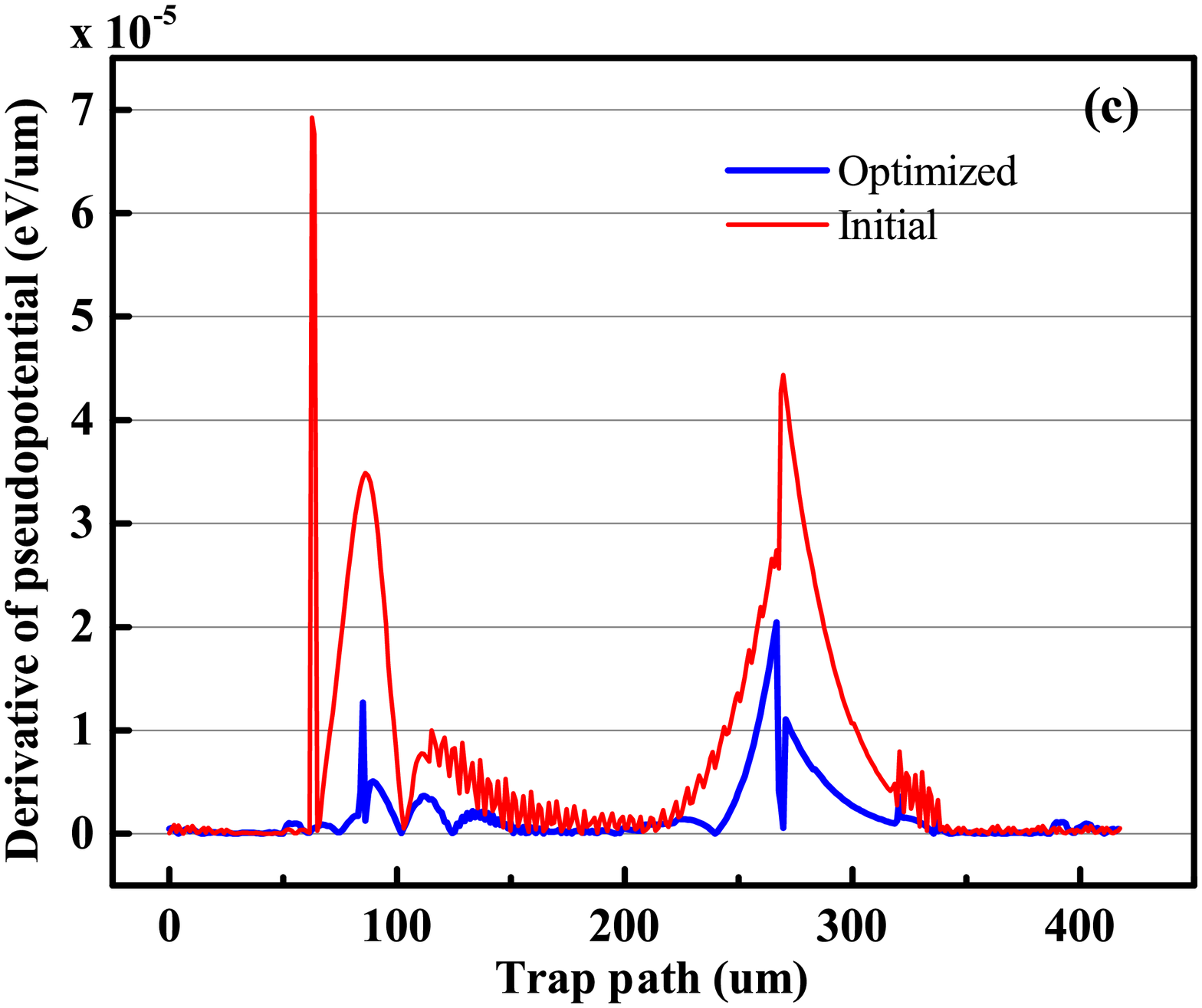}
\quad
\includegraphics[width=7.55cm,height=6.3cm]{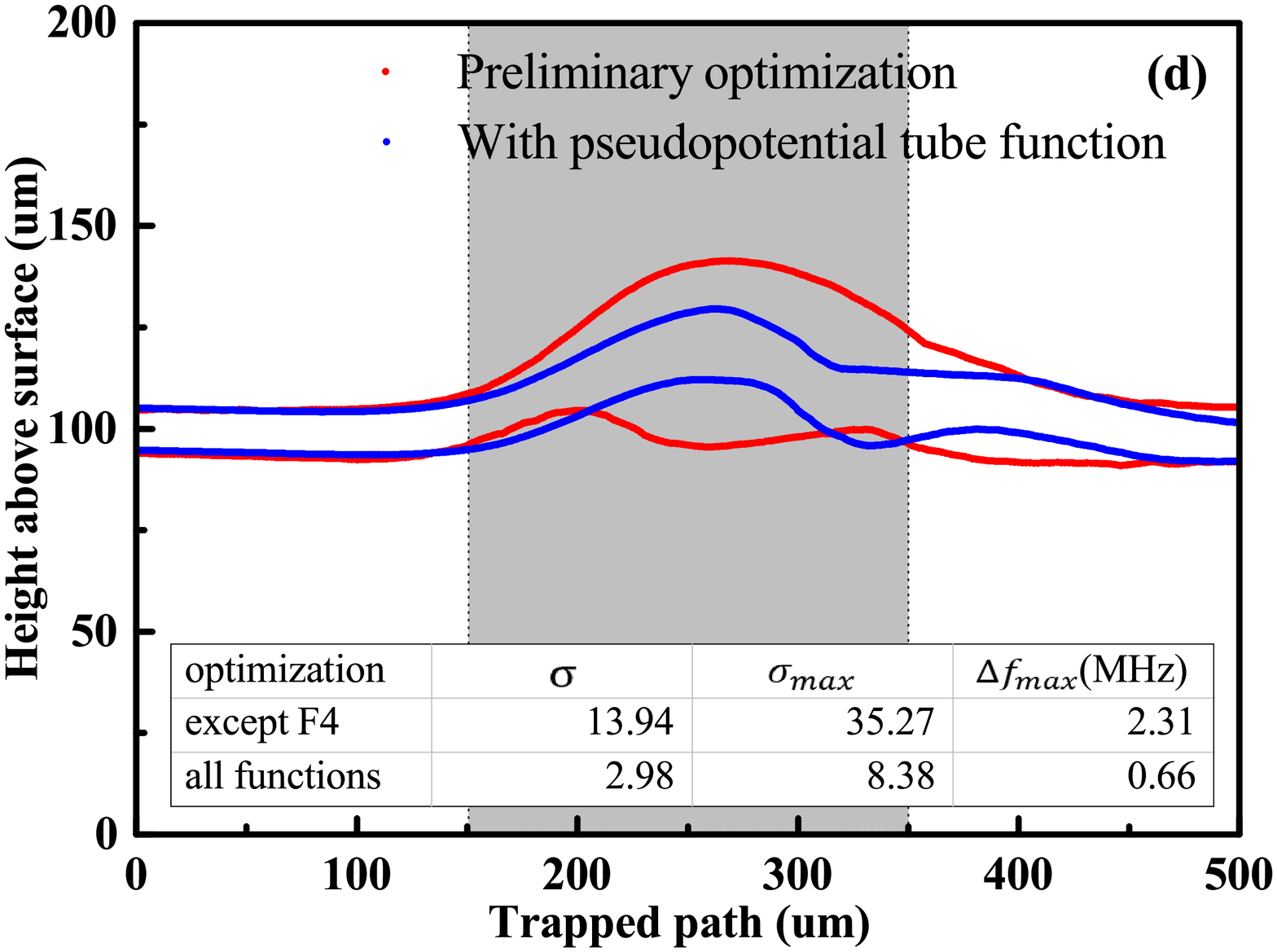}
 \caption{\label{vid:PRSTPER.4.010101}%
Results of the multi-objective optimization for the junction (red and blue lines denote initial and optimized results). (a) Fluctuation of trapped height. (b) Pseudopotential barriers of the initial and optimized junction.  (c) Derivative of the pseudopotential barriers in (b).  (d) Scale ($y$ direction) of the RF pseudopotential tube at 10 meV from the RF null line. Here, $\sigma$ and $\Delta\sigma$ are the standard deviation and the maximum deviation of the pseudopotential tube radius in the shuttling path. $\Delta f_{max}$ is the secular-frequency difference between the junction zone and linear zone.
 }%
\end{figure*}
In the optimal process, the value of the multi-objective function in Eq. (5) decreases from 219 to 5.39 after just one iteration, which shows that our function exhibits rapid convergence. Figures 5(a)--(d) show the results before and after optimization. In the initial junction, the trapped height fluctuates approximately 65 $\mu$m. After the optimization, the height fluctuation has been reduced by a factor of five to just 10.4 $\mu$m. The maximum pseudopotential barrier is approximately 1 meV before the optimization, which is about $0.7\%$ of the trap depth. The largest barrier is about 0.221 meV after the optimization, which is not only less than a quarter of the non-optimized value, but also three orders of magnitude smaller than the trap depth. The maximum derivative of the pseudopotential along the shuttling path is $2.15\times10^{-5}$ eV/$\mu$m, as shown in Figure 5(c), which is approximately three times smaller than the initial value. These optimizations will drastically reduce the heating rate.

The comparison of the optimization with and without function $F_4$ is shown in Figure 5(d). The gray area represents the pseudopotential distribution above the junction and the other areas are the pseudopotential distribution above the linear electrodes. The effect of unifying the shape of the pseudopotential tube is obvious from comparing the results optimized by all multi-function equations and without applying $F_4$. The standard deviation and maximum deviation of the tube radius are extremely different with and without $F_4$, giving $\sigma=2.97, \Delta\sigma_{max}=8,38$ and $\sigma=13.94, \Delta\sigma_{max}=35.27$, respectively. The secular frequency is simulated for trapping $^{40}Ca^{+}$. The maximum secular-frequency differences $\Delta f_{max}$ are -2.31 MHz (without $F_4$) and -0.66 MHz (with $F_4$) in the transportation zone, which are lower than in linear zones. These results illustrate the importance of $F_4$ for optimization.

To validate the optimized effect, the results before and after optimization are compared by FEM simulations. For trapping $^{40}Ca^{+}$ with 100 V RF source at 25 MHz frequency, the corresponding pseudopotential profiles of the initial and optimized junction are shown in Figure 6. The pseudopotential distribution on the confinement surface (which is based on the shuttling path stretching along the $y$ direction perpendicular to the trap surface) is significantly meliorated after optimization. The results are present in the form of contour lines, and it is clear that the pseudopotential generated by the initial junction suffers a discontinuity along the shuttling path, as shown in Figure 6(a). In contrast, the pseudopotential distribution after optimization is more continuous and smooth (see Figure 6(b). The optimized electrodes are significantly better by comparing the two single-value contour lines (the black lines display the pseudopotential $\Phi=10$ meV in the figure). A large potential barrier shown in Figure 6(a) is generated by the initial junction, which is unable to realize the stable trapping and shuttling ions. However, a homogeneous shape of pseudopotential tube is achieved by optimization, enabling ion shuttling with a relative low heating rate.

\begin{figure}
\includegraphics[totalheight=3.7cm]{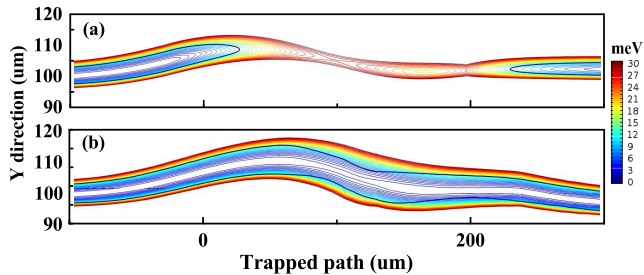}
 \caption{\label{vid:PRSTPER.4.010101}%
Pseudopotential distribution above the junction. The black solid line denotes the pseudopotential value of 10 meV. (a) and (b) are contour maps of the pseudopotential distribution generated by the initial and the optimized junction.
 }%
\end{figure}

\section{Conclusion}

\quad \quad A versatile surface ion trap is proposed which concludes a SW trap, a FW trap, and a fork junction for bridging different quantum zones. The novel SW trap of the surface ion trap enables the generation of double wells with innately rotating principle axes. Our proposed design can be used to scaling up trapped-ion-based quantum system because a pair of parallel ion chains can be trapped at a distance of 200 $\mu$m. Moreover, the distance between the double wells can be adjusted by controlling the balance of the RF voltages applied to the $RF_{Centre}$ and $RF_{Side}$ electrodes\cite{Tanaka2014Design}. A symmetric FW trap for loading ions can alleviate the pollution of the other zones during the loading process, and a fork junction optimized by a multi-objective function is used to transport ions between different zones with a low heating rate. Furthermore, our trap can be considered as an ion mixer, which can generate a mixed and ordered ion chain from the two-species ions trapped in the double wells. This is expected to be useful in sympathetic cooling scheme. The trap can also be applied as a beam splitter for electrons, which is very useful in electronic imaging technology\cite{hammer2015microwave}. We can also use the trap to rapidly split ion chains, like in an electron beam splitter.

The trap not only solves technical problems in cooling experiments by rotated principle-axes, but also provides an effective way to increase the number of trapped ion quantum bits. In our design, ions no longer exist as a single linear chain in the radial $x$ direction\cite{Ou2016Optimization,Xie2017Creating}, but as multiple ion chains spread over the trapped surface, forming a quantum network. In future studies, it should be possible to increase the number of RF electrodes, such as in interdigital electrodes\cite{Zhang2016Sniffing}, to produce many parallel ion chains over the trap surface. There will be spin-spin interaction between the different ion trains in the axial direction. By controlling the distance between the ion chains, it may be possible to store and exchange information, as well as apply quantum logic control between adjacent ion chains. The results of our study may provide additional theoretical concepts and experimental breakthrough for quantum computation, precision measurements, and quantum metrology.

\section*{Acknowledgments}

\quad \quad We would like to thank Wei Liu for fruitful discussions. This work is supported by the National Basic Research Program of China under Grant No. 2016YFA0301903; the National Natural Science Foundation of China under Grant Nos. 11174370, 11304387, 61632021, 11305262, 61205108, and 11574398, and the Research Plan Project of the National University of Defense Technology under Grant Nos. ZK18-03-04 and ZK16-03-04.

\section*{References}

\bibliographystyle{iopart-num}
\bibliography{IOPLaTeXGuidelines}

\end{document}